\documentclass[twocolumn,floatfix,prl]{revtex4-2}%
\usepackage{graphicx}
\usepackage{amsmath,amsfonts,amssymb}
\usepackage{mathrsfs}
\usepackage{color}
\usepackage{bm}
\def\a{{ {\alpha} }}
\def\k{{ {\bm k} }}
\def\p{{ {\bm p} }}
\def\q{{ {\bm q} }}
\def\Q{{ {\bm Q} }}

\def\0{{ {\bm 0} }}
\def\r{{ {\bm r} }}

\def\s{{\sigma}}
\def\e{{\epsilon}}
\allowdisplaybreaks[4]

\begin{document} 
\title{
Low temperature phase transitions inside CDW phase in kagome metals AV$_3$Sb$_5$ (A=Cs,Rb,K): Significance of mixed-type Fermi surface electron correlations
}
\author{
Jianxin Huang$^1$, Rina Tazai$^2$, Youichi Yamakawa$^1$, 
Seiichiro Onari$^1$, and Hiroshi Kontani$^1$}

\date{\today }

\begin{abstract}
To understand the multistage phase transitions in V-based kagome metals
inside the charge-density-wave (CDW) phase,
we focus on the impact of the "mixed-type" Fermi surface
because it is intact in the CDW state on the "pure-type" Fermi surface.
On the mixed-type Fermi surface, moderate 
spin correlations develop, and we reveal that uniform ($\q={\bm0}$) bond order
is caused by the paramagnon interference mechanism,
which is described by the Aslamazov-Larkin vertex correction.
A dominant solution is the $E_{2g}$-symmetry nematic order,
in which the director can be rotated arbitrarily.
In addition, we obtain the $A_{1g}$-symmetry order,
which leads to the change in the lattice constants without symmetry breaking. The predicted $E_{2g}$ and $A_{1g}$ channel fluctuations at $\q={\bm0}$ can be observed by the elastoresistance measurements.
These results are useful to understand the multistage phase
transitions inside the $2\times2$ CDW phase.
The present theory has a general significance because mixed-type Fermi Surfaces (with multiorbital van-Hove singularities)
exist in various kagome lattice systems.

\end{abstract}

\affiliation{
$^1$Department of Physics, Nagoya University,
Nagoya 464-8602, Japan
\\
$^2$Yukawa Institute for Theoretical Physics, Kyoto University,
Kyoto 606-8502, Japan 
}
 
\sloppy

\maketitle

\begin{figure}[htb]
\includegraphics[width=.9\linewidth]{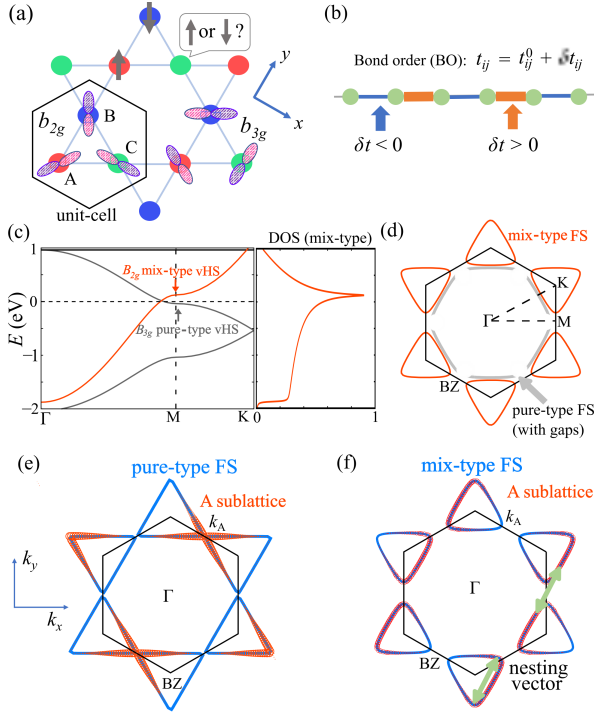}
\caption{
(a) Kagome lattice structure.
$b_{3g}$- and $b_{2g}$-orbitals are shown.
(b) BO parameter given by the hopping integral modulation $\delta t_{ij}$.
(c) $b_{3g}$-orbital (with pure-type vHS) and $b_{2g}$-orbital (with mixed-type vHS) band structure the DOS.
(d) FSs of pure-type FS ($n^{\rm pure}=2.6$) 
and mixed-type FS ($n^{\rm mix}=1.6$).
Here, the pure-type FS exhibits the $2\times2$ BO induced gap.
(e) A-sublattice weight shown by red color on the pure-type FS at 
$n_{\rm vHF}^{\rm pure}=2.5$.
The vHS point at $\k=\k_{\rm A}$ is composed of A sublattice.
(f) A-sublattice weight on the mixed-type FS
at $n_{\rm vHF}^{\rm mix}=1.6$.
The vHS point at $\k=\k_{\rm A}$ is composed of B+C sublattices.
}
\label{fig:fig1}
\end{figure}

{\it Introduction}---
Exotic electronic states and correlation-driven superconductivity 
in kagome metals AV$_3$Sb$_5$ (A=K, Cs, and Rb)
have attracted increasing attention.
Strong Coulomb interaction and the geometrical frustration of V site
electrons (in Fig. \ref{fig:fig1}(a))
give rise to exotic quantum phase transition without magnetization. 
At ambient pressure ($P=0$), AV$_3$Sb$_5$ exhibits $2\times2$ 
bond-order (BO) 
at $T_{\rm BO}=78, 94$ and $102$ K for A=K, Cs and Rb, respectively
\cite{kagome-exp1,kagome-exp2,NMR1,STM1,STM2}.
The BO is the correlation-driven modulation of the hopping integrals
($\delta t_{ij}$) (Fig. \ref{fig:fig1}(b)). 
The superconducting (SC) state appears at $T_c=1\sim3K$ inside the BO phase
\cite{SC1,SC}.
Under pressure, $T_{\rm BO}$ gradually decreases,
while $T_c$ exhibits non-monotonic pressure dependence
\cite{Pressure-dep}.
The maximum $T_c \ (\sim10{\rm K})$ is realized 
around the BO critical pressure $P_c^{\rm BO}\approx 2$GPa,
consistently with the BO fluctuation pairing 
mechanism proposed in Ref. \cite{Tazai-kagome}.
The predicted $s$-wave superconductivity has been recently confirmed
by the penetration depth and electron irradiation measurements
\cite{Roppongi,SC2}.


Rich symmetry-breaking states 
"inside the BO phase" have been a significant open problem.
For example, the $C_6$ symmetry-breaking nematic state 
\cite{elastoresistance-kagome,birefringence-kagome}
and the time-reversal-symmetry breaking (TRSB)
without spin order
\cite{STM1,muSR3-Cs,muSR2-K,muSR4-Cs,muSR5-Rb,eMChA}
have been reported.
These states are caused by
the correlation-driven hopping integral modulation:
$\delta t_{ij}$.
The nematic order is given by the
bond order with real $\delta t_{ij}$,
and the TRSB order is given by the imaginary $\delta t_{ij}$.
The latter accompanies topological charge-current
\cite{Haldane}
that gives the giant anomalous Hall effect (AHE)
\cite{AHE1,AHE2}.
Theoretically, sizable off-site (beyond-mean-field) interaction
\cite{Tazai-kagome,Onari-SCVC,Yamakawa-FeSe,Onari-FeSe,Onari-TBG,Tsuchiizu1,Tsuchiizu4,Chubukov-PRX2016,Fernandes-rev2018,Kontani-AdvPhys,Tazai-rev,Kawaguchi,Yamakawa-cuprate,Davis-rev2013,Thomale2021,Neupert2021,Balents2021,Tazai-kagome2,Tazai-kagome-GL,PDW-kagome,Fernandes-CLC,Thomale-GL,Nat-g-ology,email,Tazai-HF,Tazai-CeB6}
leads to non-local order parameters ($\delta t_{ij}\ne0$) in kagome metals.

At present, the possible onset temperatures 
of symmetry-breaking states "inside the BO phase" are unsolved.
One possible state, the TRSB order parameter strongly develops below $T^* = 35\sim50$K 
in the $\mu$-SR \cite{eMChA,muSR3-Cs,muSR4-Cs,muSR5-Rb}
and the AHE \cite{AHE1,AHE2} measurements.
As for another possibility, the nematic state, 
the scanning birefringence study \cite{birefringence-kagome}
reports $T_{\rm nem}\approx T_{\rm BO}$,
while $T_{\rm nem}\approx 35$K is reported by 
the elastoresistance measurement
\cite{elastoresistance-kagome}.
This discrepancy may indicate that 
the BO layer stacking with $\pi$-shift leads to 
the weak nematicity at $T_{\rm BO}$,
and another nematic order emerges inside the BO phase.
The origin of the latter nematicity is not understood at present.

In V-based kagome metals, two major Fermi surfaces (FSs) are 
composed of the $b_{3g}$-orbitals and the $b_{2g}$-orbitals
(Fig. \ref{fig:fig1}(a)).
The bandstructure and the FSs are shown in 
Figs. \ref{fig:fig1}(c) and (d), respectively.
Both FSs give large density-of-states (DOS) at the Fermi level
due to the van-Hove singularity (vHS) at three M points.
Near the Fermi level, the $b_{3g}$-orbital FS is called the "pure-FS",
where each vHS point is composed of a single sublattice;
see Fig. \ref{fig:fig1}(e).
The vHS points are gapped by the $2\times2$ BO on the pure-FS below $T_{\rm BO}$;
see the unfolded FS in Fig. \ref{fig:fig1}(d).
Previous theories have mainly been devoted to understand 
the significant roles of the pure-type FS
\cite{Thomale2021,Neupert2021,Balents2021,Tazai-kagome}
except for Refs.
\cite{Nat-g-ology,Fernandes-CLC}.
However, one may expect that an additional phase transition
below $T_{\rm BO}$ would occur on the $b_{2g}$-orbital FS, the mixed-type FS
because it is not harmed by the BO.
Its sublattice weight is shown in Fig. \ref{fig:fig1}(f),
In addition, it is notable that the mixed-type FS universally exists in usual kagome lattice models.
Therefore, the research on the mixed-type FS is of great significance.

%

In this paper,
to understand the phase transitions below $T_{\rm BO}$ in AV$_3$Sb$_5$,
we focus on the impact of the mixed-type FS 
because it is intact in the BO phase.
On the mixed-type FS, moderate antiferromagnetic (AFM)
spin correlations develop, which leads to the uniform ($\q={\bm0}$) bond order by the paramagnon interference mechanism.
This beyond-mean-field mechanism is described by the 
Aslamazov-Larkin (AL) vertex corrections.
A dominant BO solution is the $E_{2g}$-symmetry nematic order
in which the director can be rotated arbitrarily.
In addition, we obtain the $A_{1g}$-symmetry bond order which would accompany the change in the lattice constants.
These results are useful to understand the multistage phase
transitions inside the
$2\times2$ BO phase.

{\it Model Hamiltonian and Formulations.}---
In this letter, we study the bond-order phase transition mediated by the quantum fluctuations
in kagome lattice Hubbard model with a "mixed-type" FS 
composed of three $b_{2g}$ orbitals. 
We apply the density-wave (DW) equation method 
to derive the optimized order parameter 
(= symmetry-breaking self-energy $\Delta\Sigma$)
driven by the beyond-mean-field vertex corrections.
In Ref. \cite{Tazai-kagome},
the present authors studied the $b_{2g}$+$b_{3g}$ orbital
kagome lattice Hubbard model for AV$_3$Sb$_5$,
and found that the pure-type FS alone gives the $2\times2$ BO state.
However, the $b_{2g}$ orbital FS can induce
different instabilities for $T\ll T_{\rm BO}$.
For this reason, to find the phase transition inside the BO phase,
we study the $b_{2g}$ orbital kagome lattice tight-binding Hubbard model.

The $b_{2g}$ orbitals at sublattices A-C are shown in Fig. \ref{fig:fig1}(a).
The corresponding mixed-type FS is shown in Fig. \ref{fig:fig1}(d).
The kinetic term of the $b_{2g}$ orbital model is 
$H_0=\sum_{\k,l,m,\s} h_{lm}(\k) c^{\dagger}_{\k,l,\sigma} c_{\k,m,\sigma}$,
where $h_{lm}(\k)$ is the 
momentum representation of the nearest-neighbor hopping integral $t_{lm}(r)$, $l,m={\rm A, B, C}$ and $\s$ is the spin index.
In our study the unit of energy (Coulomb interaction, hopping integral and temperature) is eV. 
We set the nearest-neighbor hopping integral $t=-0.5$
\cite{Thomale2021,Tazai-kagome}
and put the number of electrons on the mixed-type FS $n^{\rm mix}=1.6$. 
The $3\times3$ Green function is given as
${\hat G}(\k,\e_n)=[(i\e_n+\mu){\hat 1}-{\hat h}(\k)]^{-1}$,
where $\e_n=(2n+1)\pi T$ is the fermion Matsubara frequency.
We also introduce the on-site Coulomb interaction term 
$H_U= U\sum_{i,l}n_{i,l,\uparrow}n_{i,l,\downarrow}$,
where $n_{i,l,\s}$ is the electron number at unit cell $i$.

\begin{figure}[htb]
\includegraphics[width=.8\linewidth]{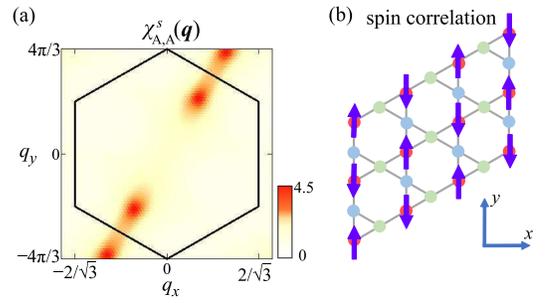}
\caption{
(a) Obtained $\chi^s_{\rm A,A}(\q)$ 
that shows the AFM correlation.
(b) Real-space short-range spin correlation.}
\label{fig:fig2}
\end{figure}

In the mean-field-level approximation, the spin instability is the most prominent \cite{Tazai-kagome}. 
In the random phase approximation (RPA), spin susceptibility is 
$\hat{\chi}^s(q)=\hat{\chi}^0(q)(\hat{1}-U\hat{\chi}^0(q))^{-1}$,
where $\chi^0_{l,m}(q)=-T\sum_{k}G_{l,m}(k+q)G_{m,l}(k)$ is the $3\times3$ irreducible susceptibility matrix
and $q\equiv(\q,\omega_l=2\pi Tl)$. 
The spin susceptibility diverges when the spin Stoner factor $\alpha_s$,
which is defined as the maximum eigenvalue of $U\hat{\chi}^0(q)$, reaches unity.
Figure \ref{fig:fig2}(a) shows the spin susceptibility $\chi^s_{\rm A,A}(\q)$
at $T=0.02$ and $\alpha_s=0.97$ ($U=2.2$).
Its maximum peak appears at the nesting of the A-sublattice FS 
shown in Fig. \ref{fig:fig1}(f). 
The real-space short-range spin correlation is depicted in Fig. \ref{fig:fig2}(b). In contrast, such AFM spin fluctuations remain small in the pure-type case. The  obtained $\chi^s(\q)$ is used to construct the kernel function of the DW equation, $I_q^{L,M}(k,p)$ in Eq. (\ref{eqn:DWeq}), which is a non-linear function of $\chi^s(\q)$. We will show that the AFM spin fluctuations in the mixed-type FS give rise to the exotic ferro-bond orders via the AL vertex corrections.

\begin{figure}[htb]
\includegraphics[width=.99\linewidth]{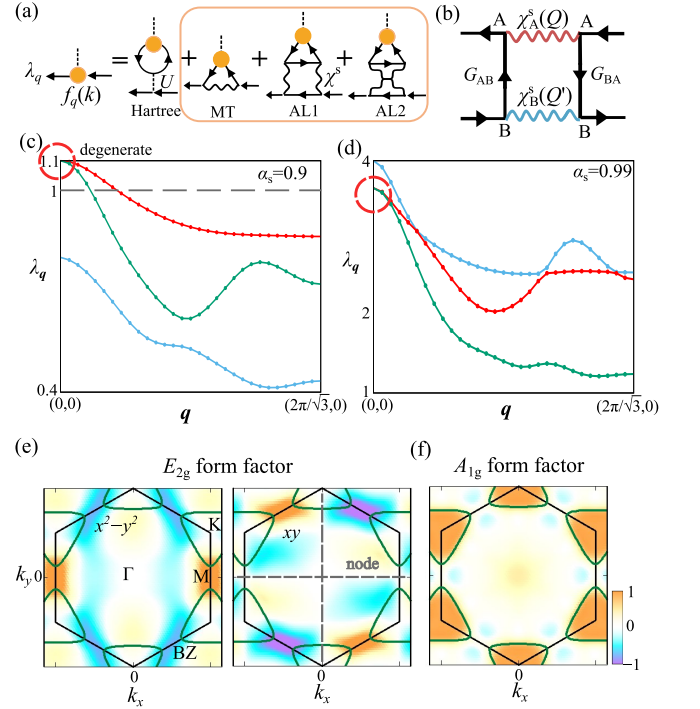}
\caption{
(a) Linearized DW equation with respect to the 
form factor $f_\q(k)$ and eigenvalue $\lambda_\q$.
The p-h pairing interaction is composed of the Hartree term,
the MT term, and the AL terms. The solid line is the non-interacting Green function, and the wave lines are the spin fluctuations from RPA.
(b) Interference between two paramagnons (${\bm Q}$ and ${\bm Q}'$) 
that leads to the DW order at $\q={\bm Q}-{\bm Q}'$.
(c-d) Eigenvalue $\lambda_\q$ for two cases: 
(c) Moderate ($\a_S=0.9$) and 
(d) strong ($\a_S=0.99$) spin fluctuation.
Three states are shown in different colors. 
In both cases, $\lambda_\q$ exhibits the maximum at $\q={\bm0}$.
(e-f) Diagonal form factor $\sum_{l}^{\rm A,B,C}f_{\q={\bm0}}^{ll}(k)$:
(e) $E_{2g}$ symmetry solution obtained for $\a_S=0.9$ and 
(f) $A_{1g}$ symmetry solution obtained for $\a_S=0.99$. 
Note that off-diagonal form factor is also large.
}
\label{fig:fig3}
\end{figure}

{\it Density Wave equation analysis.}---
The nonmagnetic DW orders cannot be obtained by RPA in the Hubbard model with the on-site Coulomb interaction, because the charge Stoner factor $\alpha_c$ is always smaller than spin Stoner factor $\alpha_s$ in mean-field approximation. However, the vertex corrections in the DW equation (Fig \ref{fig:fig3}(a)) give various charge-channel DW states due to the "paramagnon interference processes". 

From a field-theoretical point of view, the nonmagnetic bond-order is described as the symmetry-breaking self-energy $\Delta\Sigma$, which is nothing but the DW order parameter \cite{Kontani-AdvPhys}.
($\Delta\Sigma$ is much smaller than the self-energy without symmetry-breaking $\Sigma_0$.)
Once $\Delta\Sigma$ emerges, it will give the symmetry-breaking in $\chi^0(q)=-T\sum_k G(k+q)G(k)$, where $G$ is the Green function with the total self-energy $\Sigma_{tot}(k)=\Delta\Sigma(k)+\Sigma_{0}(k)$.
In the meanwhile, when symmetry-breaking in $\chi^s(\q)$ emerges, 
the self-energy $\Sigma(k)=\frac{3U^2}{2}T\sum_q\chi^s(q)G(k+q)$
will contain finite $\Delta\Sigma= \Sigma(k)-\Sigma_{0}(k)$. 
Such feedback between $\Delta\Sigma$ and $\chi^s(\q)$ becomes positive when $\lambda>1$, so the nonmagnetic bond-order occurs spontaneously. In the DW equation, the form factor describes $\Delta\Sigma$, and the AL terms give the positive feedback effect \cite{Tazai-LW}.
This theory was first developed to explain the orbital nematic state
in Fe-based superconductors,
and it has been applied to the cuprates and the twisted-bilayer graphene successfully
\cite{Kontani-AdvPhys}.
Recently, the BO and current ordered states in kagome metals with pure-type band also have been explained by the AL processes \cite{Tazai-kagome,Tazai-kagome2}.

The DW equation for the charge-channel order is
\begin{eqnarray}
\lambda_{\q}f_\q^{L}(k)&=& -\frac{T}{N}\sum_{p,M_1,M_2}
I_\q^{L,M_1}(k,p) 
\nonumber \\
& &\times \{ G(p)G(p+\q) \}^{M_1,M_2} f_\q^{M_2}(p) ,
\label{eqn:DWeq}
\end{eqnarray}
where $I_\q^{L,M}(k,p)$ is the "particle-hole (p-h) pairing interaction", the kernel function for charge channel. The detail of it is in the SM A \cite{SM}.
$k\equiv (\k,\e_n)$ and $p\equiv (\p,\e_m)$ ($\e_n$, $\e_m$ are fermion Matsubara frequencies). 
$L\equiv (l,l')$ and $M_i\equiv(m_i,m_i')$ 
represent the pair of sublattice indices A, B, C. 
$\lambda_{\q}$ is the eigenvalue that represents the instability of the DW at wavevector $\q$, 
and $\max_\q\{\lambda_\q\}$ reaches unity at $T=T_{\rm DW}$. 
$f_\q^L(k)$ is the Hermitian form factor that is proportional to the p-h condensation 
$\sum_\s \{ \langle c_{\k+\q,l,\s}^\dagger c_{\k,l',\s}\rangle - \langle \cdots\rangle_0 \}$, 
or equivalently, 
the symmetry-breaking self-energy $\Delta\Sigma$. 
In real space, the $\k$-dependent form factor gives the correlated hopping between $(i,l)$ and $(j,m)$,
which is given as:
$\delta t_{il,jm} (=\delta t_{jm,il})^{*}$.
%
\begin{eqnarray}
\delta t_{il,jm} &=& \frac1N \sum_{\k}f_{\q}^{lm}(\k) e^{i\k\cdot(\r_{il}-\r_{jm})}e^{i\q\cdot r_{il}} ,
\label{eqn:form-r}
\end{eqnarray}
where $\r_{il}$ represents position of $l$ sublattice in the unit-cell $i$.
The BO preserves the time-reversal-symmetry: $\delta t_{il,jm}=\delta t_{jm,il}=\mathrm{real}$.

The kernel function $I_\q^{L,M}$ is given by the 
functional derivative of the Luttinger ward function $\Phi_{\rm{LW}}(G)$.
Here, we apply the fluctuation-exchange approximation for $\Phi_{\rm{LW}}(G)$.
Then, $I_\q^{L,M}$ is composed of one Hartree term, one single-magnon exchange Maki-Thompson (MT) term 
and two double-magnon interference Aslamazov-Larkin (AL) terms, as depicted in Fig. \ref{fig:fig3}(a)
\cite{Kontani-AdvPhys}. 
Importantly, the AL terms represent the "interference between two paramagnons with momenta ${\bm Q}$ and ${\bm Q}'$"
that leads to the DW order at $\q={\bm Q}-{\bm Q}'$;
see Fig. \ref{fig:fig3}(b).

With the obtained spin susceptibility as the paramagnon, we solve the DW equation. 
The $\q$-dependence of the eigenvalue $\lambda_\q$ at $n^{\rm mix}=1.6$ ($T=0.02$) 
is exhibited in Figs. \ref{fig:fig3}(c) and (d).
(Here, we show $\lambda_\q$ only on the path $\Gamma$-$M$
because $\lambda_\q$ on the path $M$-$K$-$\Gamma$ is small.)
Figure \ref{fig:fig3}(c) [(d)] shows the result for $\alpha_s=0.9 \ (0.99)$,
where spin fluctuation strength is moderate (strong).
Importantly, the maximum peak position is located at $\q={\bm0}$ in both cases.
The obtained $\q={\bm0}$ order,
which is very different from the nesting driven $\q\ne{\rm 0}$ order,
is naturally derived from the interference between two paramagnons 
with wavevectors ${\bm Q}$ and ${\bm Q}'(={\bm Q})$
\cite{Kontani-AdvPhys}.
For $\a_S=0.9$ in Fig. \ref{fig:fig3} (c),
the largest two eigenvalues are degenerate, 
and the corresponding form factors are $E_{2g}$ symmetry BO;
(${\hat f}_{x^2-y^2}(k)$, ${\hat f}_{xy}(k)$).
The corresponding diagonal form factor $\sum_l f_{\q={\bm0}}^{ll}(k)$ 
is shown in Fig. \ref{fig:fig3}(e).
Note that off-diagonal form factors are as large as the diagonal form factors.
For $\a_S=0.99$ in Fig. \ref{fig:fig3}(d),
the non-degenerate largest eigenvalue
corresponds to $A_{1g}$ symmetry form factor shown in Fig. \ref{fig:fig3}(f).

\begin{figure}[htb]
\includegraphics[width=.99\linewidth]{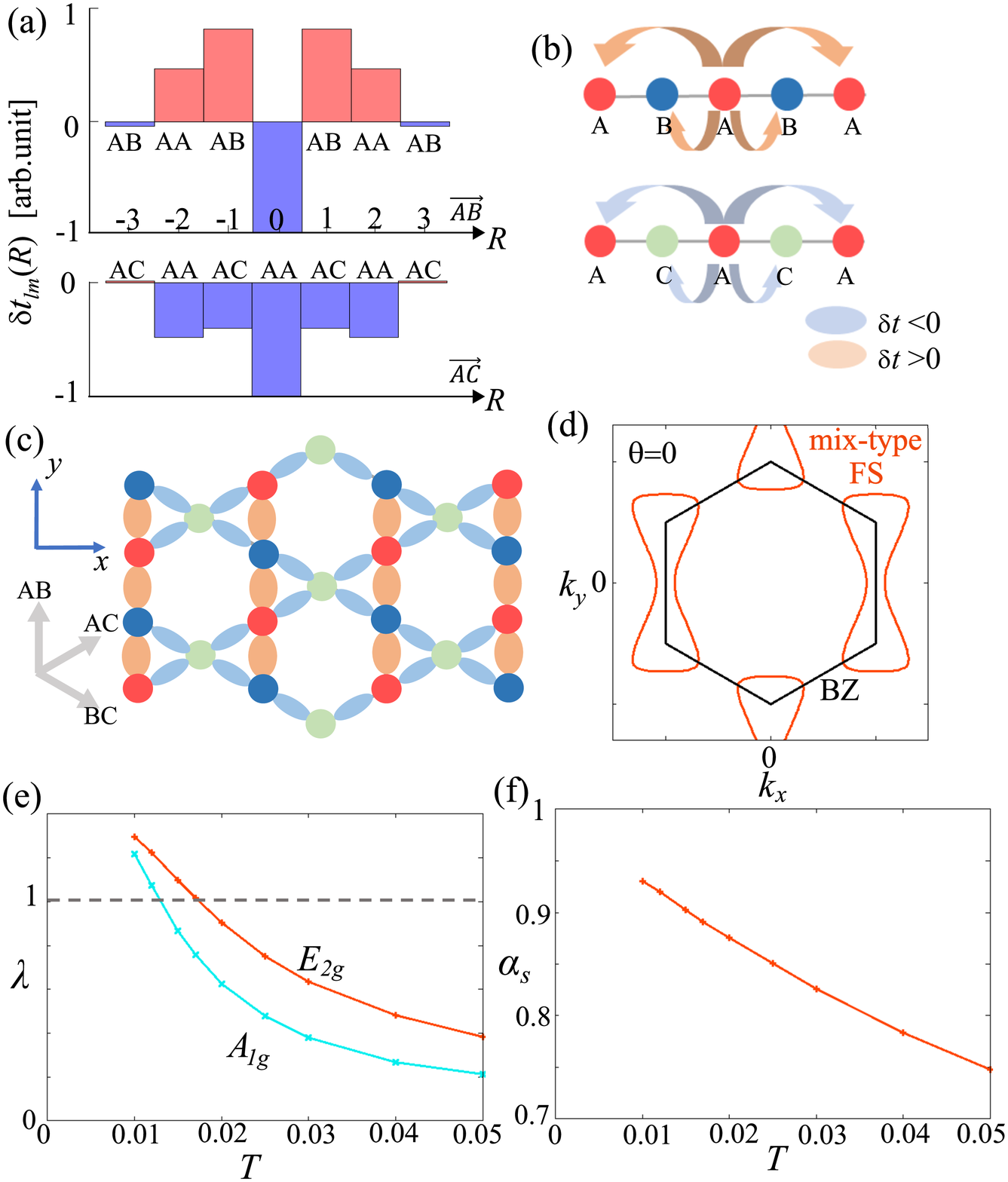}
\caption{
(a) Obtained $\delta t_{lm}(R)$ along two directions. 
The horizontal axis means the real space distance. 
For example, in AB direction, $R=1 \ (2)$ marks the hopping modulation 
between the nearest A and B (A and A).
Note that $\delta t_{\rm AC}(R)=\delta t_{\rm BC}(R)$.
(b) Schematic pictures of $\delta t_{lm}(R)$.
(c) Nematic bond order in real space for ${\hat f}_{x^2-y^2}$. 
Orange color represents $\delta t_{ij}>0$,
while blue color represents $\delta t_{ij}<0$.
(d) Nematic FS induced by the nematic BO ${\hat f}_\theta$. 
It can be rotated to any angle by changing the angle $\theta$.
(e) $T$-dependence of $\lambda_{\q=\0}$ for $E_{2g}$ and $A_{1g}$ states,
for $U=1.95$.
(f) $T$-dependence of $\a_S$.
}
\label{fig:fig4}
\end{figure}

The $E_{2g}$ symmetry BO gives the nematicity, 
and its director can be rotated at any angle by taking 
the linear combination of ${\hat f}_{x^2-y^2}(k)$ and ${\hat f}_{xy}(k)$
\cite{Onari-TBG}.
To understand the nature of the $E_{2g}$ symmetry nematic state,
we discuss the real-space form factor $\delta t_{il,jm}$ based on Eq. (\ref{eqn:form-r}).
When the wavevector is $\q={\bm0}$,
$\delta t_{il,jm}$ is simply written as 
$\delta t_{lm}({\bm r})$ with ${\bm r}\equiv {\bm r}_{il}-{\bm r}_{jm}$.
The obtained $\delta t_{lm}({\bm r})$ for $f^{lm}_{x^2-y^2}(k)$ at $\q={\bm0}$ 
is shown in Fig. \ref{fig:fig4}(a).
(Its schematic picture is given in Fig. \ref{fig:fig4}(b).)
It is plotted in two directions (AB and AC) from sublattice A,
as a function of the distance $R$, where the distance between nearest sites is 1. 
The BO along BC direction is the same as that along AC direction. 
(The even-parity relation $\delta t_{lm}(R)=\delta t_{ml}(-R)$ is verified.)
$\delta t_{ll}$ at $R=0$ represents the onsite charge modulation at sublattice $l$,
and $\delta t_{lm}$ at $R=\pm1$ ($l\ne m$) represents BO between the nearest sites.

A schematic picture of the hopping modulation due to the nearest BO
is given in Fig. \ref{fig:fig4}(c).
In this case, the lattice structure will become nematic
in the presence of finite electron-phonon coupling 
due to $E_{2g}$ acoustic mode.
We also discuss the FS deformation due to the $E_{2g}$ nematic BO,
under the order parameter 
$H'=\Delta E \sum_{\k\s} {\hat c}_{\k\s}^\dagger{\hat f}_\theta(\k){\hat c}_{\k\s}$, where 
${\hat f}_\theta(\k) \equiv \cos\theta {\hat f}_{x^2-y^2}(\k) + \sin\theta {\hat f}_{xy}(\k)$.
The ordered nematic FS at $\theta=0$ 
with $\Delta E=0.01$
is showed in Fig. \ref{fig:fig4}(d). 
By changing the angle $\theta$, the nematic FS can be rotated to any direction.

Another state, the $A_{1g}$ BO, does not break the symmetry of the system. However, it induces the FS deformation due to the hopping modulation and the change in the lattice constants should be accompanied through the finite electron-phonon coupling. 
In SM B \cite{SM}, we present the nontrivial hopping modulation caused by the $A_{1g}$ BO. 

Figure \ref{fig:fig4}(e) shows the 
$T$-dependence of $\lambda$ for $E_{2g}$ and $A_{1g}$ states,
in the case of $U=1.95$.
The obtained $\lambda$ monotonically increases with decreasing $T$,
while the increment of $\a_S$ is moderate (see Fig. \ref{fig:fig4}(f)).
Although the obtained transition temperature is relatively high,
it will be reduced to below $T_{\rm BO}$ by introducing the self-energy effect
\cite{Tazai-LW}. 
In fact, due to the 
ordinary self-energy $\Sigma_0$ which breaks no symmetry, the Green function is given as
$G=1/(i\epsilon_n-\Sigma_0-\epsilon_k)\approx z/(i\epsilon_n-z\epsilon_k)$, 
where the mass-enhancement factor $z^{-1}=m^*/m$ is larger than 1.
The factor $z$ leads to the rescaling, $T\to zT$ and $U\to U/z$, under the condition of the fixed $\alpha$ and $\lambda$\cite{Yamakawa-FeSe}. Thus, the transition temperature will be reduced by $z$. 
It is an important future issue to derive a reliable transition temperature by taking account of the self-energy.

In the SM C \cite{SM}, 
we discuss the important roles of the AL and MT vertex corrections
for the BO.
We show that the AL term gives large positive contribution
for both $E_{2g}$ and $A_{1g}$ BOs.
In addition, it is found that the MT term slightly favors the nodal $A_{1g}$ BO.
Interestingly, the same AL process leads to the $\q={\bm0}$ BO in mixed-type FS model and $2\times2$ BO in the pure-type FS model.

Finally, we discuss the results of the existing elastoresistance experiments \cite{elastoresistance2,elastoresistance3,elastoresistance4} based on the present theory. In Ref. \cite{elastoresistance2}, significant $E_{2g}$ symmetry channel nematic correlation is observed under $T_{\rm CDW}$ and above a transition temperature $T_{\rm nem} \sim 35K$. In contrast, in other experiments\cite{elastoresistance3,elastoresistance4}, strong enhancement of $A_{1g}$ channel susceptibility is reported below $T_{\rm CDW}$ and presents a phase transition at $T^*\sim 20K$. Importantly, both  $E_{2g}$ and $A_{1g}$ instabilities develop in the mixed-type FS in the present theory, and the leading instability can exchange. Thus, different experimental results may originate from a slight sample dependence. Noteworthy, Ref. \cite{Moll-strain} reports the nematic order is strongly stabilized under tiny (extrinsic) strain in kagome metal. Thus, the DW equation analysis for the “distorted kagome metal model” would be useful to understand the variety of the experimental reports. This is an important future issue.

{\it Summary.} ---
We studied the phase transitions below $T_{\rm BO}$ in AV$_3$Sb$_5$
by focusing on the mixed-type FS that is intact in the BO phase.
It is revealed that the paramagnon interference mechanism 
leads to the uniform ($\q={\bm0}$) bond order.
A dominant solution is the $E_{2g}$-symmetry nematic order.
We also obtain the $A_{1g}$-symmetry non-nematic order.
These results are useful to understand the multistage phase transitions
inside the
$2\times2$ BO phase. 
This theory has a general significance because mixed-type like multi-sublattice FS exists in many kagome metals, such as ATi$_3$Bi$_5$ \cite{Ti-based kagome}. In this case, the intra-sublattice nesting gives rise to the AFM spin correlation and such correlation leads to the paramagnon interference. Thus, due to this mechanism, exotic bond orders would occur in various kagome metals.

\acknowledgements
This study has been supported by Grants-in-Aid for Scientific
Research from Ministry of Education, Culture, Sports, Science and Technology  of Japan (Grants No. JP18H01175, No. JP20K03858, No. JP20K22328,
No. JP22K14003, No. JP23K03299),
and by the Quantum Liquid Crystal
No. JP19H05825 KAKENHI on Innovative Areas from Japan Society for the promotion of Science of Japan.


\clearpage
\newpage


\makeatletter
\renewcommand{\thefigure}{S\arabic{figure}}
\renewcommand{\theequation}{S\arabic{equation}}
\makeatother
\setcounter{figure}{0}
\setcounter{equation}{0}
\setcounter{page}{1}
\setcounter{section}{1}

\begin{widetext}
\begin{center}
{
\bf \large 
[Supplementary Materials]

Low temperature phase transitions inside CDW phase in kagome metals AV$_3$Sb$_5$ (A=Cs,Rb,K): Significance of mixed-type Fermi surface electron correlations

\vspace{3mm}
}
\end{center}

\begin{center}
Huang, Jianxin$^{1}$, Rina Tazai$^{2}$, Youichi Yamakawa$^{1}$,
Seiichiro Onari$^{1}$, 
and Hiroshi Kontani$^1$
\end{center}

\begin{center}
\textit{
$^1$Department of Physics, Nagoya University,
Nagoya 464-8602, Japan
\\
$^2$Yukawa Institute for Theoretical Physics, Kyoto University,
Kyoto 606-8502, Japan 
}
\end{center}

\end{widetext}
\subsection{A: Formula of DW equation}

We are going to give the formula of the kernel function of the DW equation here.
In Eq. (\ref{eqn:DWeq}), the kernel function $I^{L,M}_{\q}$ is the irreducible four-point vertex. When $\q=0$, $I^{L,M}_{\q}$ is given by the Ward identity: $I^{L,M}_{\q}=-\delta \Sigma^L(k)/\delta G^M(k')$, where $L\equiv (l,l')$, $M_i\equiv(m_i,m_i')$ represent the pair of sublattice indices A, B, C. $\Sigma$ is one-loop fluctuation exchange self-energy. The vertex correction consists of one Hartree term, one single-magnon exchange MT term and two double-magnon interference AL1 and AL2 terms. Their analytic expressions are given as

\begin{align}
&I^{l,l',m,m'}_{\q}(k,k')=\Gamma^c_{l,l',m,m'} \nonumber\\
&+\sum_{b=s,c}\frac{a^b}2 [V^b_{l,m;l',m'}(k-k')\nonumber\\
&-T\sum_p\sum_{l_1,l_2,m_1,m_2}V^b_{l,l_1;m,m_2}(p_+)V^b_{m',m_2;l',l_2}(p_-)\nonumber\\
&\times G_{l_1,l_2}(k-p)G_{m_2,m_1}(k'-p)\nonumber\\
&-T\sum_p\sum_{l_1,l_2,m_1,m_2}V^b_{l,l_1;m_2,m'}(p_+)V^b_{m_1,m;l',l_2}(p_-)\nonumber\\
&\times G_{l_1,l_2}(k-p)G_{m_2,m_1}(k'+p)]
\label{eqn:kernel function} ,
\end{align}

where the double-counting in the second order terms should be subtracted.
Here $a^{s(c)}=3(1),p_{\pm}\equiv p+\q/2,p=(p,\omega_l)$, 
and $\hat V^b$ is the $b$-channel interaction matrix given by $\hat V^b=\hat\Gamma^b+\hat\Gamma^b\hat\chi^b\hat\Gamma^b$. 
$\hat\Gamma^b$ is the $b$ channel bare multiorbital Coulomb interaction. 
(In the present model, $(\hat\Gamma)^s_{ll'mm'}=U\delta_{ll'}\delta_{l'm}\delta_{mm'}$ and $(\hat\Gamma)^c_{ll'mm'}=-U\delta_{ll'}\delta_{l'm}\delta_{mm'}$.) The susceptibility ${\hat \chi}^b(q)$ is given in the main text.
The first term in Eq. (\ref{eqn:kernel function}) is the MT term when the second and third terms are the AL terms. For the higher order term, in the functional-renormalization-group (fRG) study they have been proved unbiased to the DW equation. \cite{Tsuchiizu1,Tsuchiizu4}


\subsection{B: $A_{1g}$ mode}

In the main text, we studied the possible phase transitions in kagome lattice Hubbard model with mixed-type FS by the DW equation. 
It was found that $E_{2g}$ and $A_{1g}$ symmetry BOs can appear respectively under different spin Stoner factor $\alpha_s$.
The obtained $E_{2g}$ BO form factor and its schematic picture are exhibited in Fig \ref{fig:fig4}. 
In contrast, the $A_{1g}$ mode breaks no symmetry. Here we explain the obtained  $A_{1g}$ symmetry form factor and the corresponding schematic picture.
We choose $\alpha_s=0.975$ ($U=2.21$ and $T=0.02$), which is reported in SM B \cite{SM} as the transition point for two kinds of BOs.

\begin{figure}[htb]
\includegraphics[width=.7\linewidth]{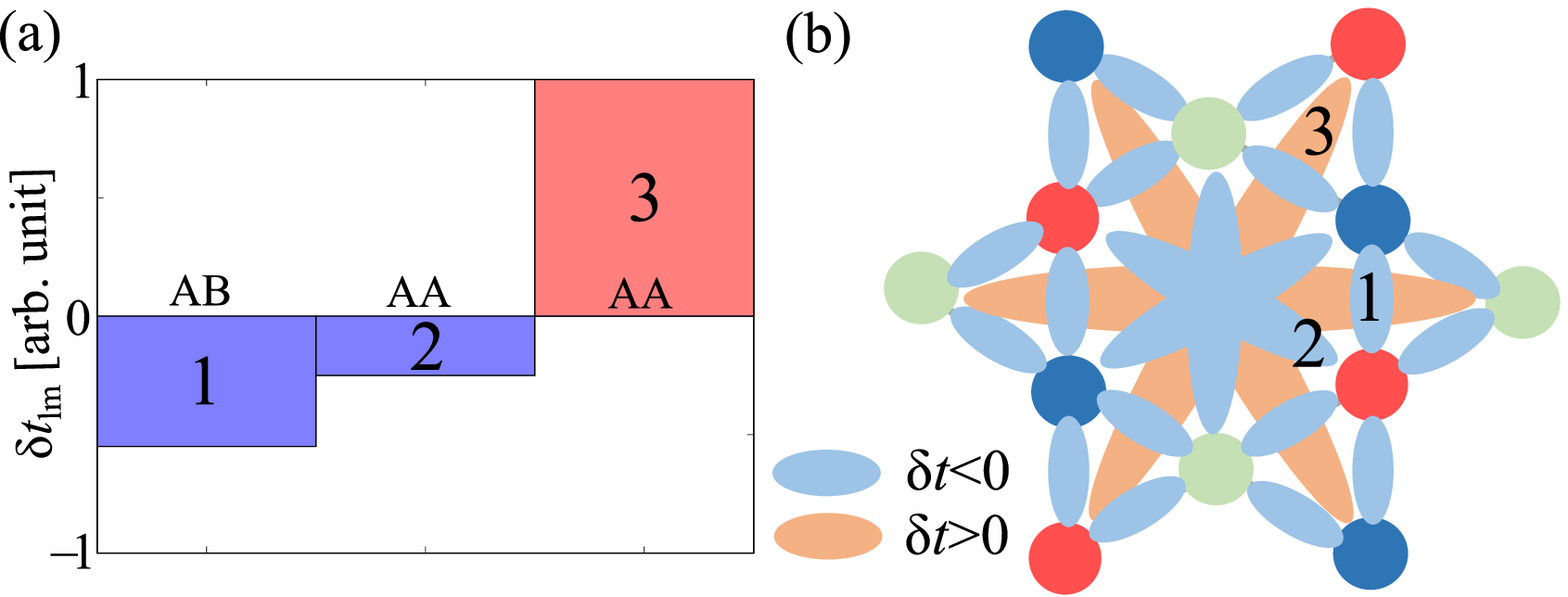}
\caption{
(a) Main hopping modulations $\delta t_{lm}$ for $A_{1g}$ mode 
($\alpha_s=0.975, T=0.02$). 
The corresponding hopping is in (b). Only hopping modulations related to sublattice $A$ are shown because $A_{1g}$ keeps symmetry. The nearest neighbor hopping and two long-range hoppings are exhibited. 
(b) Schematic picture of $A_{1g}$ BO, showing two repulsive hoppings and a long-range strong attractive one. The value of the hopping modulation for different number marks are shown in (a) respectively.
}
\label{fig:fig5}1
\end{figure}

As known from Fig. \ref{fig:fig3}(d), the $A_{1g}$ state's maximum eigenvalue also located at $\q=0$. Therefore, the real-space $A_{1g}$ form factor $\delta t_{il,jm}$ becomes intra-unit-cell, like Fig. \ref{fig:fig4} for the $E_{2g}$ order. The obtained $\delta t_{lm}$ is exhibited in Fig. \ref{fig:fig5}(a) with the three main values.
Fig. \ref{fig:fig5}(a) is the obtained numerical results, showing one nearest neighbor (bond 1) hopping modulation  and two long-range (bonds 2 and 3) hopping modulations; see Fig. \ref{fig:fig5}(b).
The schematic picture of the obtained A1g order is shown in Fig. \ref{fig:fig5}(b).
It can be seen the nearest neighbor hopping reduce. Importantly, long-range hopping modulations $\delta t_{lm}$ among bonds 2 and 3 emerge irrespective of the fact that the original $t_{lm}$ is zero. These two bonds have opposite sign. Interestingly, the obtained $\delta t_{lm}$ exhibits the sign-change depending on the distance, and $|\delta t_{lm}|$ is the largest for the next-nearest bonds. These opposite modulations of nearest neighbor hopping and long-range hoppings compete with each others.

Furthermore, the $A_{1g}$ BO will induce the $A_{1g}$ deformation of the FS due to the additional hopping $2$ and $3$ in Fig. \ref{fig:fig5}(b). In addition, an isotropic lattice deformation will be induced below the $A_{1g}$ BO transition temperature $T^{A1g}_{BO}$ through the finite electron-phonon coupling since the minimum condition of the free energy ($F_{\rm electron}+F_{\rm lattice}+F_{\rm electron-lattice}$) changes.
Importantly, the present theory naturally explains the enhancement of the $A_{1g}$ channel susceptibility observed in Ref. \cite{elastoresistance3} above 
$T_{\rm BO}^{\rm A_{1g}}$

\subsection{C: Important Roles of the AL and MT Vertex Corrections and the Quantum Interference Mechanism}

\begin{figure}[htb]
\includegraphics[width=.7\linewidth]{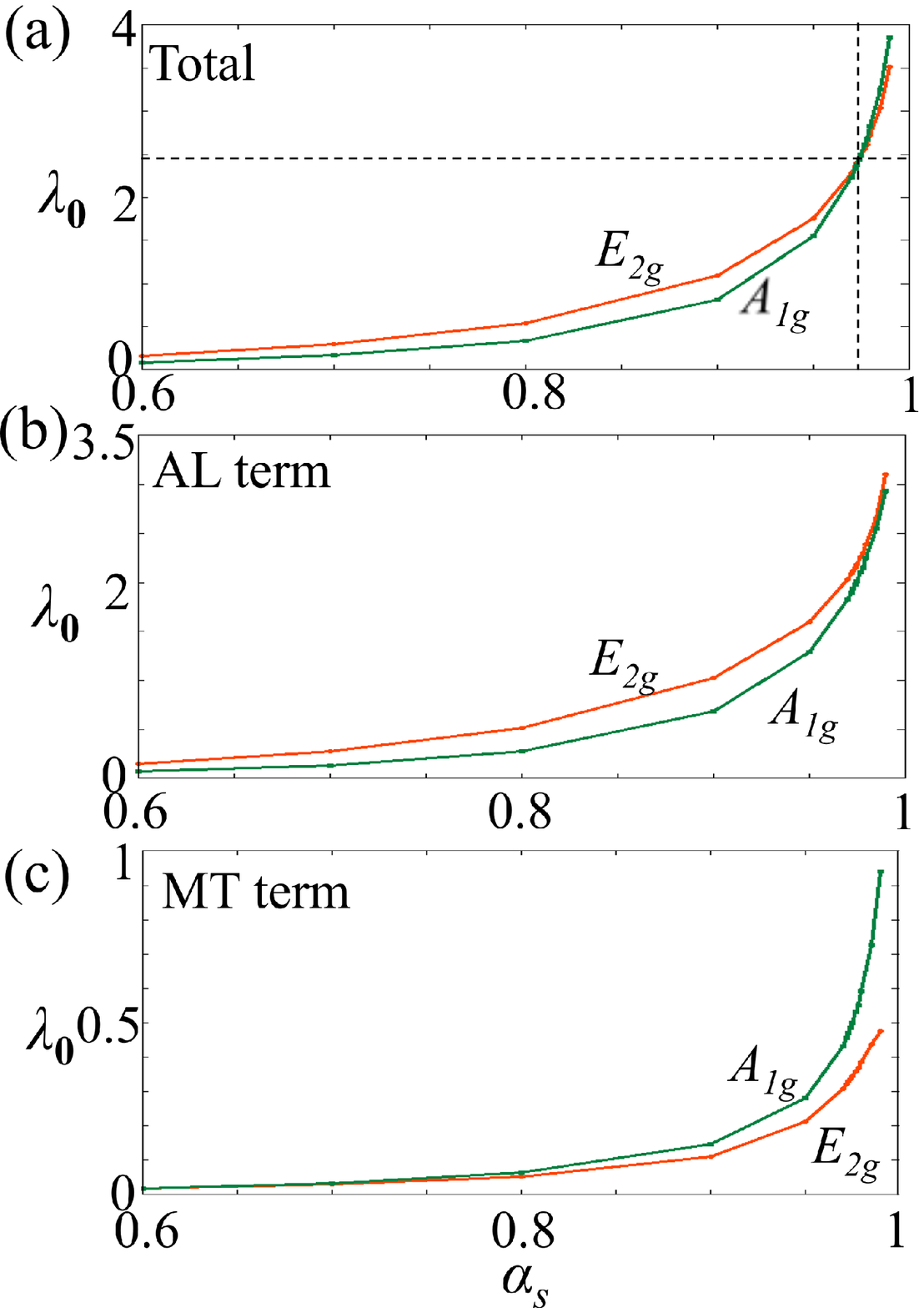}
\caption{
(a) $\lambda_{\q={\rm0}}$ for the $E_{2g}$ solution and that for the $A_{1g}$ solution
as functions of $\alpha_s$ for $T=0.02$.
These eigenvalues reach unity for $\a_S\sim0.9$.
With $\alpha_s\lesssim0.975$, the $E_{2g}$ solution has the largest $\lambda$. 
For $\alpha_s\gtrsim0.975$, it is replaced with the $A_{1g}$ solution.
(b)(c) Eigenvalue contributed from the AL and MT terms,
$\lambda_{\q}^{\rm AL}$ and $\lambda_{\q}^{\rm MT}$, respectively.
(Note that $\lambda_{\q}=\lambda_{\q}^{\rm AL}+\lambda_{\q}^{\rm MT}$.)
While the AL term gives the dominant contribution to $\lambda_{\q={\bm0}}$,
the MT term plays an important role in determining the symmetry and 
nodal structure of the form factor.
}
\label{fig:fig6}
\end{figure}

Here, we discuss the important roles of the AL and MT vertex corrections
for the nematic order.
Figure \ref{fig:fig6} (a) shows the obtained 
$\lambda_{\q={\rm0}}$ for the $E_{2g}$ solution and the $A_{1g}$ solution
as functions of $\alpha_s$ for $T=0.02$.
These eigenvalues reach unity for $\a_S\sim0.9$.
With increasing $\a_S$,
the largest eigenvalue state changes from $E_{2g}$ BO to $A_{1g}$ BO
at $\alpha_s\sim 0.975$.
(Simple charge order is prohibited by the Hartree term.)

Figures \ref{fig:fig6} (b) and (c) give the 
eigenvalue contributed from the AL and MT terms for the fixed form factor,
$\lambda_{\q}^{\rm AL}$ and $\lambda_{\q}^{\rm MT}$, respectively.
(Note that $\lambda_{\q}=\lambda_{\q}^{\rm AL}+\lambda_{\q}^{\rm MT}$.)
For both $E_{2g}$ and $A_{1g}$ channels,
the AL term gives large positive contribution.
In fact, the AL term can stabilize general even-parity BOs,
whose wavevector $\q=\Q-\Q'$ (see Fig. \ref{fig:fig3} (b) in the main text)
becomes zero or small (minor) nesting vector
\cite{Kontani-AdvPhysS}.
The AL processes give the relation 
$\lambda_{E_{2g}}^{\rm AL}\gtrsim\lambda_{A_{1g}}^{\rm AL}$;
see Fig. \ref{fig:fig6} (b).
Thus, the nodal $A_{1g}$ BO originates from the 
large MT term; see Fig. \ref{fig:fig6} (c).
The $A_{1g}$ BO with $\lambda_{\q={\bm0}}\sim1$
will be realized if the self-energy effect reduces 
the eigenvalue\cite{Kontani-AdvPhysS}.

Finally, we are going to discuss the different BO arose by mixed- and pure-type FS.
The quantum interference mechanism produces the nearest site BO 
in both the mixed-type FS and the pure-type FS models. 
However, the wavevector is $\q=\bm{0}$ in the mixed-type FS and 
$\q\ne\bm{0}$ in the pure-type FS.
To understand this difference, we focus on the significant role of the 
vHS points. 
Since the vHS is composed of two sublattices in the mixed-type FS, 
the $\q=\bm{0}$ nearest site BO is efficiently realized by 
the intra-vHS scattering. 
In contrast, in the pure-type FS, the vHS is composed of a 
single sublattice, so the inter-site BO should be given by 
the inter-vHS process at $\q=\q_{\rm vHS1}-\q_{\rm vHS2}$. 
Thus, we can understand the different BOs between both FSs 
by using the unique present theory.


\end{document}